\documentclass[aip,jcp,reprint]{revtex4-2}
\usepackage[english]{babel}
\usepackage{amsmath,amsthm,latexsym,amssymb,amsfonts,epsfig}
\usepackage{bm}
\usepackage{dsfont}
\usepackage[left = 1.35cm, right = 1.35cm, top = 2.2cm, bottom = 2.2cm]{geometry}

\usepackage{graphicx}

\usepackage{color}

\definecolor{OliveGreen}{rgb}{0.1, 0.4, 0.1}
\definecolor{caribbeangreen}{rgb}{0.0, 0.8, 0.6}
\definecolor{carminered}{rgb}{1.0, 0.0, 0.22}

\usepackage{setspace}
\emergencystretch=\maxdimen
\def\XXint#1#2#3{{\setbox0=\hbox{$#1{#2#3}{\int}$}
     \vcenter{\hbox{$#2#3$}}\kern-.5\wd0}}

\allowdisplaybreaks

\usepackage[colorlinks=true,citecolor=OliveGreen,linkcolor=OliveGreen,urlcolor=OliveGreen]{hyperref}

\usepackage{mathtools}
\usepackage{mathrsfs}

\usepackage{type1ec} %

\usepackage{amsbsy}

\newcommand{\bnabla}{\boldsymbol{\nabla}}
\newcommand{\ch}{\operatorname{ch}}
\newcommand{\sh}{\operatorname{sh}}

\newcommand{\csh}{\operatorname{csh}} 
\newcommand{\sch}{\operatorname{sch}} 

\newcommand{\e}{\hat{\bm{e}}}

\begin{document}

\title{Elastic displacements and viscous flows in wedge-shaped geometries with a straight edge: Green's functions for perpendicular forces}

\author{Abdallah Daddi-Moussa-Ider}
\email{admi2@open.ac.uk}
\thanks{corresponding author}
\affiliation{School of Mathematics and Statistics, The Open University, Walton Hall, Milton Keynes MK7 6AA, United Kingdom}

\author{Andreas M. Menzel}
\email{a.menzel@ovgu.de}

\affiliation{Institut f\"ur Physik, Otto-von-Guericke-Universit\"at Magdeburg, Universit\"atsplatz 2, 39106 Magdeburg, Germany}

\begin{abstract}
Edges are abundant when elastic solids glide in guiding rails or fluids are contained in vessels. We here address induced displacements in elastic solids or small-scale flows in viscous fluids in the vicinity of one such edge. For this purpose, we solve the governing elasticity equations for linearly elastic, potentially compressible solids, as well as the low-Reynolds-number flow equations for incompressible fluids. Technically speaking, we derive the associated Green's functions under confinement by two planar boundaries that meet at a straight edge. The two boundaries both feature no-slip or free-slip conditions, or one of these two conditions per boundary. Previously, we solved the simpler case of the force being oriented parallel to the straight edge. Here, we complement this solution by the more challenging case of the force pointing into a direction perpendicular to the edge. Together, these two cases provide the general solution. Specific situations in which our analysis may find application in terms of quantitative theoretical descriptions are particle motion in confined colloidal suspensions, dynamics of active microswimmers near edges, or actuated distortions of elastic materials due to activated contained functionalized particles. 
\end{abstract}

\maketitle

\section{Introduction}
\label{sec:intro}

In the simplest three-dimensional (3D) situation, elastic deformations are considered for linearly elastic, homogeneous, isotropic, possibly compressible bulk solids. Particularly, the associated Green's function for the displacement field as described by the associated linear Navier--Cauchy equations is available \cite{cauchy1828exercises, love2013treatise}. As a specific application, the displacements of actuated particles in linearly elastic solids can be quantified through theoretical analysis \cite{huang2016buckling, puljiz2016forces, puljiz2019displacement}.

Formally as a subclass of the Navier--Cauchy equations, the dynamics of 3D incompressible, isotropic, homogeneous bulk viscous fluids at small scales and low speeds is well understood in terms of low-Reynolds-number hydrodynamics~\cite{happel12}. The Green's function is termed Oseen tensor for the corresponding linear Stokes equations \cite{dhont96, kim13}. In these situations, the motion of passive particles in liquid suspensions or active self-propelled microswimmers in fluids are well analyzed and quantitatively described \cite{lauga09, elgeti15, lauga2016bacterial, zottl2016emergent}.

When boundaries come into play, the situation rapidly complicates. In the presence of one planar, infinitely extended no- or free-slip boundary, corresponding Green's functions are still available both for the linearly elastic solid case and the viscous fluid~\cite{blake1971note, felderhof05, menzel2017force}. Likewise, there are strategies of analytical or quasi-analytical solutions in the presence of two parallel, infinitely extended no- or free-slip boundaries \cite{liron76,felderhof06twoMem, felderhof10echoing, swan2011hydrodynamics, mathijssen2016hotspots, mathijssen2016hydrodynamics, daddi18jpcm, lutz2024internal}.
Other geometries include three-dimensionally bounded domains, which are finite systems and have been intensely studied analytically, such as spherical systems~\cite{walpole2002elastic, daddi2017hydrodynamic, daddi2017hydrodynamicII, daddi2018creeping, fischer2019magnetostriction, hoell2019creeping, sprenger2020towards, kree2021dynamics, fischer2020towards, fischer2024magnetic, kawakami2025migration}.

In the present work, we consider a 3D geometry in which two infinitely extended planar boundaries meet at a finite angle. There, they bound the system by a straight edge. 
Overall, the boundaries form a wedge-shaped confinement. 
Problems involving corner geometries have garnered considerable interest in recent decades.
In this case, a fundamentally different mathematical approach is required once the planar boundaries are not parallel to each other.
Previously, the general Stokes flow near a two-dimensional corner has been investigated~\cite{dean1949steady, moffatt1964viscous, crowdy2017analytical}.
It was observed that if one or both of the corner boundaries are no-slip walls and the corner angle is below a certain critical value, the flow near the corner exhibits a series of eddies that diminish in size and intensity with proximity to the corner.

Low-Reynolds-number flows in incompressible viscous fluids near 3D corners have been analyzed in pioneering works by Sano and Hasimoto through the study of increasingly complex wedge-shaped geometries~\cite{sano76, sano1977slow, sano1978effect, hasimoto80, sano1977slow_thesis}.
Not long ago, the 3D Stokes flow problem near a corner has been revisited and explored using a complex analysis approach\cite{dauparas2018leading, dauparas2018stokes}.
The associated Green's function under free-slip wedge-shaped confinement was recently derived for commensurate opening angles of the wedge. For this purpose, an image method was applied~\cite{sprenger2023microswimming}.
The Stokes flow in a semi-infinite wedge, bounded not only by sidewalls but also by a bottom wall, has been further investigated~\cite{shankar2000stokes}.

More recently, in our prior article~\cite{daddi2025elastic}, we addressed the more general geometry of commensurate or incommensurate opening angles of the wedge. We considered both linearly elastic solids and low-Reynolds-number fluid flows confined by the wedge, where no-slip or free-slip conditions prevailed on both boundaries or one such condition on each boundary. 
This situation is severely more challenging. Therefore, we so far had only solved the simpler case of an external point force applied to the elastic solid or viscous fluid in a direction parallel to the edge of the wedge \cite{daddi2025elastic}. By point force we understand a constant force applied at only one point in the system. The mathematical expression for the resulting elastic displacement field or flow velocity field corresponds to the fundamental solution or Green's function. 

Now, in the present work, we complete this solution. We here derive the Green's function for point forces directed perpendicular to the straight edge of the wedge. Together, due to the linearity of the underlying equations, the general case of arbitrarily directed point forces and thus the general expression for the Green's function follows by superposition of the parallel and perpendicular cases. 

We proceed in the following way. Due to possible compressibility, the description of linearly elastic solids is generally more complex than for low-Reynolds-number flows of incompressible fluids. Therefore, we focus on the equations for elastic displacement fields. The results for viscous flows can then be obtained by setting the Poisson ratio to one half, reinterpreting the displacement field as the flow field, and identifying the shear modulus of the solid with the shear viscosity of the fluid. We overview in Sec.~\ref{sec:math} the underlying mathematical framework. In Sec.~\ref{sec:greens_FKL_space}, we solve the equations after transformation into the Fourier-Kontorovich-Lebedev space. Real-space expressions for the Green's functions are provided in Sec.~\ref{sec:green_real}. 
Exact analytical solutions are derived in Sec.~\ref{sec:planar_boundary} by taking the limit of planar boundaries, thereby recovering previously known results.
We conclude the paper in Sec.~\ref{sec:concl}

\section{Mathematical formulation}
\label{sec:math}

\begin{figure}
    \centering
    \includegraphics[width=1\linewidth]{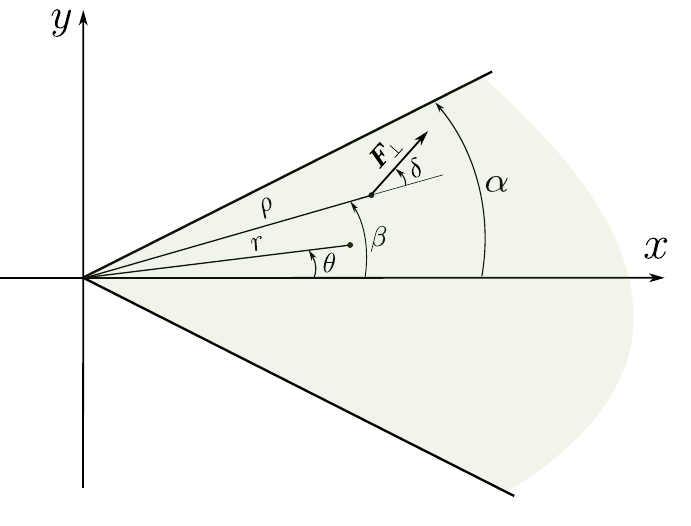}
    \caption{Schematic representation of the system setup. A wedge-shaped elastic medium with a straight edge along the $z$-axis is bounded by two planar surfaces and described in the cylindrical coordinate system \((r, \theta, z)\). The boundaries of the elastic medium are situated at polar angles \(\theta = \pm \alpha\). A point force $\bm{F}_{\perp}$ is applied perpendicular to the $z$-direction, that is, within the $x$-$y$ plane, at the location \((\rho, \beta, 0)\). $\delta$ is the angle between the force and the radial axis.}
    \label{fig:illustration}
\end{figure}

We analyze the Green's function in a wedge-shaped elastic medium with a straight edge forming its tip, as illustrated in Fig.~\ref{fig:illustration}. Our system is described in a cylindrical coordinate frame $(r, \theta, z)$. $z$-axis coincides with the straight edge of the wedge where its two surfaces meet. These two surfaces of the wedge are positioned along polar angles $\theta = \pm\alpha$, with $\alpha \in (0, \pi/2]$.
A point force is applied at $(r, \theta, z) = (\rho, \beta, 0)$, oriented perpendicular to the $z$-axis.
We define $\delta$ as the angle between the direction of the force and the radial axis. Thus, $\delta = 0$ corresponds to a purely radial force and $\delta = \pi/2$ to a purely azimuthal force.

Our objective is to determine the displacement field induced by the point force in the wedge-shaped elastic material.
The displacement field \(\bm{u}(\bm{x})\) of material elements within an isotropic, homogeneous, linearly elastic medium subjected to a force density \(\bm{f}(\bm{x})\) is governed by the Navier-Cauchy equation,
\begin{equation}
    \mu \boldsymbol{\nabla}^2 \bm{u}(\bm{x}) + \left(  \mu + \lambda \right) \boldsymbol{\nabla} \left( \boldsymbol{\nabla} \cdot \bm{u}(\bm{x}) \right) + \bm{f} (\bm{x}) = \mathbf{0} \, ,
    \label{eq:NAVI-CAUC-Eqs}
\end{equation}
with $\lambda$ and $\mu$ representing the first and second Lamé parameters, respectively, also known as the Lamé coefficients or Lamé moduli.
We define the dimensionless number 
\begin{equation}
    \sigma = \frac{ \lambda + 3\mu }{\lambda + \mu} \, , 
\end{equation}
which takes values in the interval \([1, 7)\), with the limit of incompressibility corresponding to \(\sigma = 1\).
We do not assign a direct physical interpretation to this parameter and introduce it to simplify and condense the mathematical expressions.

In the following, we examine combinations of two types of boundary conditions, no-slip (NOS) and free-slip (FRS) conditions. An NOS boundary condition indicates no relative motion between the material and the surface of the wedge.
This implies \(\bm{u} = \bm{0}\) at the boundary. An FRS boundary condition is characterized by impermeability. That is, the normal component of the displacement at the boundary, here given by \(u_\theta \), satisfies \(u_\theta = 0\), along with relaxed tangential shear, given by \(\partial u_r / \partial \theta = \partial u_z / \partial \theta = 0\).
We consider three scenarios, (i)~no-slip boundary conditions applied on both walls at \(\theta = \pm\alpha\), (ii)~free-slip boundary conditions on both walls at \(\theta = \pm\alpha\), and~(iii)~mixed boundary conditions, with NOS at \(\theta = -\alpha\) and FRS at \(\theta = \alpha\).

\subsection{Solution approach}

A general solution to Eq.~\eqref{eq:NAVI-CAUC-Eqs} is expressed as
\begin{equation}
    \bm{u} = \bnabla \left( \bm{x} \cdot \boldsymbol{\Phi} + \Phi_w \right) - (\sigma+1) \boldsymbol{\Phi} \, , 
\end{equation}
wherein $\bm{x} = (x, y, z)$ denotes the position vector in the system of Cartesian coordinates and $\boldsymbol{\Phi} = \Phi_x \, \e_x + \Phi_y \, \e_y + \Phi_z \, \e_z$.
Here, $\Phi_j$ are harmonic functions satisfying Laplace's equation, $\boldsymbol{\nabla}^2 \Phi_j = 0$, for $j \in \{x, y, z, w\}$.
Alternative representations of the general solution to the equations of equilibrium in linear elasticity have also been reported~\cite{papkovich1932solution, neuber1934neuer, naghdi1961representation, lurie2010theory, boussinesq1885application, palaniappan2011general}.

The components of the displacement field in cylindrical coordinates are given by
\begin{subequations}
\label{eq:displacement_field}
\begin{align}
    u_r &=  \left( r\, \frac{\partial}{\partial r} - \sigma \right) \Phi_r + z\, \frac{\partial \Phi_z}{\partial r} + \frac{\partial \Phi_w}{\partial r} \, , \\
    u_\theta &= \frac{\partial \Phi_r}{\partial \theta} 
    - (\sigma+1) \Phi_\theta
    + \frac{z}{r} \frac{\partial \Phi_z}{\partial \theta} + \frac{1}{r} \frac{\partial \Phi_w}{\partial \theta} \, , \\
    u_z &= r\, \frac{\partial \Phi_r}{\partial z} + 
    \left( z \, \frac{\partial}{\partial z} -\sigma \right) \Phi_z + \frac{\partial\Phi_w}{\partial z} \, ,
\end{align}
\end{subequations}
where the radial and azimuthal components of $\Phi$ are determined through coordinate transformations
\begin{equation}
    \begin{pmatrix}
        \Phi_r \\
        \Phi_\theta
    \end{pmatrix}
    =
    \begin{pmatrix}
        \cos\theta & \sin\theta \\
        -\sin\theta & \cos\theta 
    \end{pmatrix} \cdot
     \begin{pmatrix}
        \Phi_x \\
        \Phi_y
    \end{pmatrix} .
\end{equation}
In an infinitely extended, linearly elastic bulk medium, the displacement field induced by a point force~$\bm{F}$ positioned at $\bm{x}_0$ is given by~\cite{stakgold2011green}
\begin{equation}
    \bm{u}^\infty = \frac{1}{4\pi \mu \left(\sigma +1\right) s}
    \left( \sigma \, \bm{I} + \hat{\bm{s}}  \hat{\bm{s}} \right) \cdot \bm{F} ,
    \label{eq:displacement-infinity}
\end{equation}  
where $\bm{I}$ is the identity tensor, $\bm{s} = \bm{x} - \bm{x}_0$, with $s = |\bm{s}|$, $\hat{\bm{s}} = \bm{s}/s$, and $\hat{\bm{s}}  \hat{\bm{s}} \equiv \hat{\bm{s}} \, \otimes \, \hat{\bm{s}}$ denotes the dyadic product of two vectors.

In a wedge geometry, the general solution for the four harmonic functions is expressed as
\begin{equation}
    \Phi_j = \phi_j^\infty + \phi_j \, , 
\end{equation}
where \(\phi_j^\infty\), for \( j \in \{r,\theta,z,w\} \), are the free-space harmonic functions in an unbounded elastic medium, while \(\phi_j\) denote additional harmonic fields necessary to satisfy our boundary conditions.

\subsection{Integral transform analysis}

The Kontorovich-Lebedev transform (KL-transform) is a well-established technique for solving partial differential equations when a radial transform is appropriate.
It was introduced by Soviet mathematicians Kontorovich and Lebedev in the late 1930s to solve specific boundary-value problems~\cite{kontorovich1938one, kontorovich1939method, kontorovich1939application}.
The mathematical theory was developed later by Lebedev~\cite{lebedev1946, lebedev1949, Lebedev1966}.
Transform tables have been provided by Erdélyi \textit{et al.}~\cite[p.~75]{erdelyi1953higher}, along with various other types of integral transforms.
The transform has been widely used for various physical problems involving wedges, primarily in solving electromagnetic scattering and diffraction problems~\cite{lowndes1959application, rawlins1999diffraction, antipov2002diffraction, salem2006electromagnetic, hwang2009scattering, kim2009electromagnetic, shanin2011modified, eom2014integral, lyalinov2016integral} or studying fluid flows~\cite{waechter1969steady, waechter1969steadyB}. 
Other applications include analyzing the Schr\"odinger equation \cite{yakubovich2013use, lyalinov2020functional} and quantifying the escape probability for integrated Brownian motion with nonzero drift~\cite{atkinson1994escape}.

In the following, we denote the hyperbolic sine and hyperbolic cosine functions by $\sh$ and $\ch$, respectively. Additionally, the hyperbolic secant and hyperbolic cosecant functions are denoted by $\sch$ and $\csh$, respectively, and are defined as $\sch(x) = 1/\ch(x)$ and $\csh(x) = 1/\sh(x)$.

Combining the KL-transform with the classical Fourier transform of the axial coordinate yields the Fourier-Kontorovich-Lebedev transform (FKL-transform), defined by the double integral over the radial and axial coordinates 
\begin{align}
\widetilde{f}(\nu, k) &=
    \mathscr{T}_{i\nu} \left\{ f(r,z) \right\} \notag \\[3pt]
    &= 
    \int_{-\infty}^\infty \mathrm{d}k \, e^{ikz}
    \int_0^\infty f(r) K_{i\nu} (|k|r) \, r^{-1} \, \mathrm{d} r \, ,
    \label{eq:FKL}
\end{align}
where \( K_{i\nu} \) denotes the modified Bessel function of the second kind~\cite{abramowitz72}, also known in the context of index transforms as Macdonald's function, with purely imaginary order \( i\nu \).
Note that for a positive argument, \( K_{i\nu} \) results in a real number when \( \nu \) is constrained to be real.
It is clear that the transform in Eq.~\eqref{eq:FKL} leaves the polar angle unaffected.
The inverse FKL-transform is given by the double integral
\begin{align}
    f(r,z) &=  
    \mathscr{T}_{i\nu}^{-1} \{ \widetilde{f}(\nu, k) \}
    = \frac{1}{\pi^3} 
    \int_{-\infty}^\infty \mathrm{d}k \, e^{-ikz} \notag \\[3pt]
    &\quad\times \int_0^\infty \widetilde{f}(\nu, k) \, K_{i\nu} (|k|r) \sh (\pi\nu) \, \nu \, \mathrm{d} \nu \, . \label{eq:inv_FKL}
\end{align} 
We here use a positive exponent for the forward Fourier transform and a negative exponent for the inverse Fourier transform.

Applying this transform, we convert the partial differential equation, Eq.~\eqref{eq:NAVI-CAUC-Eqs}, that governs the displacement field in an elastic medium into a set of ordinary differential equations for the polar angle \( \theta \).
We find that the FKL-transform of the Laplace equation results in a homogeneous second-order ordinary differential equation for the transformed function, given by 
\begin{equation}
    \left( \frac{\partial^2}{\partial\theta^2} - \nu^2 \right) \widetilde{f} = 0 \, .
    \label{eq:ODE-2nd-order}
\end{equation}
We have included the details of the derivation in the Appendix.

In the following, we utilize a key property of the FKL-transform related to division by \( r \). By applying the recurrence relation that links Macdonald's function to its adjacent orders, namely
\begin{equation}
    \frac{K_{i\nu}(|k|r)}{r} = 
    \frac{|k|}{2i\nu} \left(
    K_{i\nu+1}(|k|r) - K_{i\nu-1}(|k|r) \right) ,
\end{equation}
it can be shown that
\begin{equation}
    \mathscr{T}_{i\nu} \left\{ \frac{f}{r} \right\} = 
    \frac{|k|}{2i\nu} \, \big( \mathscr{T}_{i\nu+1} \{f\} - \mathscr{T}_{i\nu-1} \{f\} \big) \, .
    \label{eq:KL-property}
\end{equation}

\section{Green's function in FKL space}
\label{sec:greens_FKL_space}

\subsection{Green's function in an infinite medium}

In an unbounded medium, the displacement field caused by a point force applied in the direction perpendicular to the wedge is derived from
\begin{subequations}
    \begin{align}
    \phi_x^\infty &= \chi \cos(\beta+\delta) \, , \\
    \phi_y^\infty &= \chi \sin(\beta+\delta) \, , \\
    \phi_w^\infty &= -\rho \chi \cos\delta \, ,
\end{align}
\end{subequations}
and $\phi_z^\infty = 0$.
Here, we have defined
\begin{equation}
    \chi = -\frac{q_\perp}{s} \, , \qquad
    q_\perp = \frac{F_\perp}{4\pi \mu \left(1+\sigma \right)} \, , \label{eq:chi}
\end{equation}
with the distance from the singularity position
\begin{equation}
    s = \left( r^2+\rho^2-2\rho r \cos (\theta-\beta)+z^2 \right)^\frac{1}{2} \, .
    \label{eq:s}
\end{equation}
As demonstrated in Ref.~\onlinecite{daddi2025elastic}, the FKL-transform of \( \chi \) is given by
\begin{equation}
    \widetilde{\chi} =
    -2\pi q_\perp \, 
    \frac{\ch \left( ( \pi-|\theta-\beta| )\nu \right)}{\nu \sh(\pi\nu)} \, K_{i\nu} (|k|\rho) \, .
    \label{eq:chi_FKL}
\end{equation}
It follows from the coordinate transforms that the radial and azimuthal components of the free-space contribution to the displacement field are given by
\begin{equation}
    \phi_r^\infty = \chi \cos(\theta-\beta-\delta) \, , \quad
    \phi_\theta^\infty = -\chi \sin(\theta-\beta-\delta) \, .
\end{equation}

\subsection{General solution}

Since they are harmonic functions that satisfy the Laplace equation, it follows from Eq.~\eqref{eq:ODE-2nd-order} that the general solution for \( \widetilde{\phi}_j \), where \( j \in \{x,y,z,w\} \), is given by
\begin{equation}
    \widetilde{\phi}_j = A_j \sh(\theta\nu) + B_j \ch(\theta\nu) \, ,
    \label{eq:general_solution_phi_j}
\end{equation}
with \( A_j \) and \( B_j \) as unknown coefficients. They will subsequently be determined from the boundary conditions imposed at \( \theta = \pm\alpha \).

We express the coefficients in the form
\begin{subequations}
    \begin{align}
    A_j &= \frac{2\pi q_\perp}{\nu\sh(\pi\nu)} \,
    \left( \Lambda_j + \Omega_j \, \rho \, \frac{\partial}{\partial\rho} \right) K_{i\nu}(|k|\rho) \, , \\
    B_j &= \frac{2\pi q_\perp}{\nu\sh(\pi\nu)} \,
    \left( \Lambda_j^\dagger + \Omega_j^\dagger \, \rho \, \frac{\partial}{\partial\rho} \right) K_{i\nu}(|k|\rho) \, .
\end{align}
\end{subequations}
Here, \( \Lambda_j, \Lambda_j^\dagger, \Omega_j \), and $\Omega_j^\dagger$ are functions of \( \nu \) and depend solely on the parameters \( \alpha \), \( \beta \), \( \delta \), and \( \sigma \). 
We note that $\Omega_w$ and $\Omega_w^\dagger$ are non-vanishing only for \( j \in \{x, y\} \).
In the following, we provide their full expressions for the three types of boundary conditions considered in this study.

\subsection{No-slip--no-slip boundary conditions (2NOS)}

We require the displacement field to vanish at both surfaces of the wedge. Thus, we impose the condition $u_r = u_\theta = u_z = 0$ at $\theta = \pm\alpha$.
Since there are eight unknown coefficients and only six equations for the boundary conditions, we have some flexibility in choosing two of the coefficients. From the radial and axial components of the displacement field given by Eq.~\eqref{eq:displacement_field}, we select
\begin{equation}
    \phi_w + \phi_w^\infty = 0 \, , \quad
    \phi_r + \phi_r^\infty = 0 \, , \quad
    \phi_z = 0 \, ,  
\end{equation}
evaluated at $\theta = \pm\alpha$.
Since the general solution for \( \phi_z \) is given by Eq.~\eqref{eq:general_solution_phi_j}, and it must vanish at both boundaries, it follows that \( \phi_z \) is zero everywhere.

The condition for vanishing azimuthal displacement leads to
\begin{equation}
    \hspace{-0.1cm}
    \frac{\partial}{\partial\theta}
    \left( \phi_r + \phi_r^\infty 
    + \frac{\phi_w + \phi_w^\infty}{r} \right)
    - (\sigma+1) \left( \phi_\theta + \phi_\theta^\infty \right) = 0 , 
\end{equation}
evaluated at $\theta = \pm\alpha$.
Defining the abbreviation 
\begin{equation}
    \vartheta = \theta - \beta - \delta \, ,
\end{equation}
it follows that the corresponding equations in FKL-space are derived as
\begin{subequations} 
    \label{eq:BC_2NS}
    \begin{align}
    \widetilde{\phi}_w - \rho \, \widetilde{\chi} \,\cos\delta &= 0 \, ,  \label{eq:BC_2NS_W} \\[3pt] 
    \widetilde{\phi}_r + \widetilde{\chi} \, \cos\vartheta &= 0 \, ,  \label{eq:BC_2NS_R} \\[3pt]
    \frac{\partial \widetilde{\phi}_r}{\partial \theta} - (\sigma+1) \, \widetilde{\phi}_\theta 
    + \frac{\partial}{\partial\theta} \,\mathscr{T}_{i\nu} \left\{ \frac{ \widetilde{\phi}_w }{r} \right\} + \mathcal{R} &= 0 \, ,  \label{eq:BC_2NS_The}
\end{align}
\end{subequations}
evaluated at $\theta = \pm\alpha$, with the right-hand side
\begin{equation}
    \mathcal{R} = \sigma \widetilde{\chi} \, \sin\vartheta
    +\frac{\partial \widetilde{\chi}}{\partial \theta} \, \cos\vartheta-\rho \cos\delta \, 
    \frac{\partial}{\partial\theta} \,
    \mathscr{T}_{i\nu}
    \left\{ \frac{ \widetilde{\chi} }{r} \right\} .
\end{equation}

The unknown coefficients are determined by substituting the forms in Eqs.~\eqref{eq:general_solution_phi_j} into Eqs.~\eqref{eq:BC_2NS} and solving the resulting system of equations.  
For future reference, we introduce the abbreviations
$\zeta_\pm = \alpha \pm \beta$, 
$\xi_\pm = \alpha \pm \beta \pm \delta$, 
$\alpha_\pm = \pi \pm \alpha$,
$\beta_\pm = \pi \pm \beta$, and
$\tau = \beta+\delta$.

The coefficients associated with \( \widetilde{\phi}_w \) take a simple form because their governing equation is decoupled from the other fields. They are expressed as
\begin{subequations}
    \begin{align}
    \Lambda_w &= - \rho 
    \sh(\beta\nu) \sh \left( \alpha_- \nu \right) \csh(\alpha\nu) \cos\delta \, , \\
    \Lambda_w^\dagger &= - \rho 
    \ch(\beta\nu) \ch \left( \alpha_- \nu \right) \sch(\alpha\nu) \cos\delta \, .
\end{align}
\end{subequations}
In addition, $\Omega_w = \Omega_w^\dagger = 0$.
Contrariwise, the remaining coefficients associated with \( \widetilde{\phi}_x \) and \( \widetilde{\phi}_y \) have a more delicate structure and are written as 
\begin{subequations}
    \begin{align}
    \Lambda_x &= \Pi_+ \left( b \cos\tau + \nu \left( {a}_- + {c}_- \ch(\alpha\nu) \right) \sin\alpha \right) , \\[3pt]
    \Lambda_y &= \Pi_- \left( b \sin\tau + \nu \left( {a}_+ - {c}_+ \ch(\alpha\nu) \right) \cos\alpha \right) , \\[3pt]
    \Lambda_x^\dagger &= 
    \Pi_- \left( b^\dagger \cos\tau - \nu \left( {a}_+^\dagger - {c}_+ \sh(\alpha\nu) \right) \sin\alpha \right) , \\
    \Lambda_y^\dagger &= 
    \Pi_+ \left( b^\dagger \sin\tau - \nu \left( {a}_-^\dagger + {c}_- \sh(\alpha\nu) \right) \cos\alpha \right) .
\end{align}
\end{subequations}
In addition,
\begin{subequations}
    \begin{align}
    \Omega_x &= - \Pi_+ d_- \sin\alpha \ch(\alpha\nu) \, , \\[3pt]
    \Omega_y &= \phantom{+} \Pi_- d_+ \cos\alpha \ch(\alpha\nu) \, , \\[3pt]
    \Omega_x^\dagger &= -\Pi_- d_+ \sin\alpha \sh(\alpha\nu) \, , \\[3pt]
    \Omega_y^\dagger &= \phantom{+} \Pi_+ d_- \cos\alpha \sh(\alpha\nu) \, , 
\end{align}
\end{subequations}
where
\begin{equation}
    \Pi_\pm^{-1} = \sigma\sh(2\alpha\nu) \pm \nu \sin(2\alpha) \, .
\end{equation}
For convenience, we define the following expressions to present the results in a more compact form
\begin{subequations}
    \begin{align}
    {a}_\pm &=  \cos\xi_+ \sh\left( \beta_- \nu\right) \pm \cos\xi_- \sh\left( \beta_+\nu\right) \, , \\[3pt]
    {a}_\pm^\dagger &=  \cos\xi_+ \ch\left( \beta_- \nu\right) \pm \cos\xi_- \ch\left( \beta_+\nu\right) \, , 
\end{align}
\end{subequations}
together with
\begin{subequations}
    \begin{align}
        {c}_\pm &= Z_\pm \left( \cos\zeta_-\ch \left( \zeta_+\nu \right) \pm 
    \cos\zeta_+ \ch \left( \zeta_-\nu\right) \right) \, , \\[3pt]
    d_\pm &= Z_\pm \left( \sin\zeta_-\sh \left( \zeta_+\nu \right) \pm 
    \sin\zeta_+ \sh \left( \zeta_-\nu\right) \right) \, ,
    \end{align}
\end{subequations}
where $Z_\pm = 2\cos\delta \sh(\pi\nu) \, \Delta_\pm$ with
\begin{equation}
    \Delta_\pm^{-1} = \ch\left( 2\alpha\nu\right) \pm \cos(2\alpha) \, .
    \label{eq:Delta}
\end{equation}
Moreover, 
\begin{subequations}
    \begin{align}
    b &= 2\sigma \ch(\alpha\nu) \sh (\alpha_-\nu) \sh(\beta\nu) \, , \\[3pt]
    b^\dagger &= 2\sigma \sh(\alpha\nu) \ch (\alpha_-\nu) \ch(\beta\nu) \, .
    \end{align}
\end{subequations}

\subsection{Free-slip--free-slip boundary conditions (2FRS)}

Next, we impose free-slip conditions on both surfaces by requiring that
\begin{equation}
    u_\theta = 0 \, , \qquad
    \frac{\partial u_r}{\partial \theta} = 0 \, , \qquad
    \frac{\partial u_z}{\partial \theta} = 0 \, , 
\end{equation}
evaluated at $\theta = \pm\alpha$.
A simple form of the solution is obtained by imposing that
\begin{subequations}
    \begin{align}
    \frac{\partial}{\partial\theta} \left( \phi_r + \phi_r^\infty \right) &= 0 \, , \hspace{1.25cm}
    \phi_\theta + \phi_\theta^\infty = 0 \, , \\[3pt]
    \frac{\partial \phi_z}{\partial\theta} &= 0 \, , \quad
     \frac{\partial}{\partial\theta} \left( \phi_w + \phi_w^\infty \right) = 0 \, , 
\end{align}
\end{subequations}
evaluated at $\theta = \pm\alpha$.
We use the same reasoning as for 2NOS. Since the general solution for \( \phi_z \) is given by Eq.~\eqref{eq:general_solution_phi_j} and since its derivative with respect to \( \theta \) must vanish at both boundaries, \( \phi_z \) is identically zero.
The corresponding expressions in FKL-space read
\begin{subequations}
    \label{eq:BC_2FS}
    \begin{align}
    \frac{\partial \widetilde{\phi}_r }{\partial\theta} 
    - \widetilde{\chi} \, \sin \vartheta
    + \frac{ \partial \widetilde{\chi} }{\partial\theta} \, \cos\vartheta &= 0 \, , \\[3pt]
    \widetilde{\phi}_\theta - \widetilde{\chi} \, \sin \vartheta &= 0 \, , \\[3pt]
    \frac{\partial \widetilde{\phi}_w}{\partial\theta} - \rho  \, \frac{\partial \widetilde{\chi} }{\partial\theta} \, \cos\delta &= 0 \, ,
\end{align}
\end{subequations}
evaluated at $\theta = \pm\alpha$.

The coefficients of \( \widetilde{\phi}_w \) again have a simple form because its governing equation is decoupled from the other fields and can be written as
\begin{subequations}
    \begin{align}
    \Lambda_w &=  \rho  \sh(\beta\nu) \ch \left( \alpha_- \nu \right) \sch(\alpha\nu)\cos\delta \, , \\
    \Lambda_w^\dagger &= \rho \ch(\beta\nu) \sh \left( \alpha_- \nu \right)\csh(\alpha\nu)  \cos\delta \, ,
\end{align}
\end{subequations}
and $\Omega_w = \Omega_w^\dagger = 0$.
Defining 
\begin{subequations}
    \begin{align}
    W_x &= \tfrac{1}{2} \left( \cos\hat{\xi}_+ \sh(\beta_-\nu) - \cos\hat{\xi}_- \sh(\beta_+\nu) \right) , \\[3pt]
    W_x^\dagger &= \tfrac{1}{2} \left( \cos\hat{\xi}_+ \ch(\beta_-\nu) + \cos\hat{\xi}_- \ch(\beta_+\nu) \right)  , \\[3pt]
    W_y &= \tfrac{1}{2} \left( \sin\hat{\xi}_+ \sh(\beta_-\nu) + \sin\hat{\xi}_- \sh(\beta_+\nu) \right)  , \\[3pt]
    W_y^\dagger &= \tfrac{1}{2} \left( \sin\hat{\xi}_+ \ch(\beta_-\nu) - \sin\hat{\xi}_- \ch(\beta_+\nu) \right) ,
\end{align}
\end{subequations}
with $\hat{\xi}_\pm = 2\alpha \pm \beta \pm \delta$, we obtain
\begin{subequations}
    \begin{align}
    \Lambda_x &= \phantom{+} \Delta_+ \left( W_x - \cos\tau \sh(\beta\nu) \ch \left( \hat{\alpha} \nu \right) \right) , \\[3pt]
    \Lambda_x^\dagger &= -\Delta_- \left(  W_x^\dagger - \cos\tau \ch(\beta\nu) \ch \left( \hat{\alpha} \nu \right) \right) , \\[3pt]
    \Lambda_y &= -\Delta_- \left( W_y + \sin\tau \sh(\beta\nu) \ch \left( \hat{\alpha} \nu \right) \right) , \\[3pt]
     \Lambda_y^\dagger &= \phantom{+} \Delta_+ \left( W_y^\dagger + \sin\tau \ch(\beta\nu) \ch \left( \hat{\alpha} \nu \right) \right) ,
\end{align}
\end{subequations}
where $\hat{\alpha} = \pi-2\alpha$.
We recall that $\Delta_\pm$ was previously defined in Eq.~\eqref{eq:Delta}.
In addition, $\Omega_j = \Omega_j^\dagger = 0$ for $j \in \{x,y\}$.

\subsection{No-slip--free-slip boundary conditions (NOS--FRS)}

Finally, we consider the case of mixed boundary conditions by imposing an NOS condition at $\theta = -\alpha$ and an FRS condition at $\theta = \alpha$. Specifically, we enforce Eqs.~\eqref{eq:BC_2NS} at $\theta = -\alpha$ and Eqs.~\eqref{eq:BC_2FS} at $\theta = \alpha$.
The coefficients take more complex expressions due to the varied boundary conditions.
We obtain for the $w$-related coefficients
\begin{subequations}
    \begin{align}
    \Lambda_w &= \rho \, \frac{\ch(\beta\nu)\sh(\pi\nu) + \sh(\beta\nu) \ch \left( \hat{\alpha} \nu \right)}{\ch(2\alpha\nu)} \,  \cos\delta \, , \\
    \Lambda_w^\dagger &= \rho \, \frac{\sh(\beta\nu)\sh(\pi\nu) - \ch(\beta\nu) \ch \left( \hat{\alpha} \nu \right)}{\ch(2\alpha\nu)} \,  \cos\delta \, .
\end{align}
\end{subequations}
In addition, $\Omega_w = \Omega_w^\dagger = 0$.

For convenience, we now introduce the abbreviations $\hat{\zeta}_\pm = 2\alpha \pm \beta$, $\overline{\alpha} = \pi-4\alpha$, $\varsigma = 3\alpha-\beta$, and $\eta = 3\alpha-\beta-\delta$.
The remaining coefficients are cast in the form
\begin{subequations}
    \begin{align}
    \Lambda_j &= Q \big( \nu \left( Z G_j + H_j \right) + \sigma K_j \big) \, , \\[3pt]
    \Lambda_j^\dagger &= Q \left( \nu \left( Z G_j^\dagger -H_j^\dagger \right) -\sigma K_j^\dagger \right) , 
\end{align}
\end{subequations}
and 
\begin{equation}
    \Omega_j = -Q Y G_j \, , \quad
    \Omega_j^\dagger = -Q Y G_j^\dagger \, ,
\end{equation}
for $j \in \{x,y\}$, where we have defined
\begin{equation}
    Q = \left(\sigma \sh(4\alpha\nu) - \nu\sin(4\alpha)\right)^{-1} \, .
\end{equation}
The other coefficients are expressed as
\begin{subequations}
    \begin{align}
    Z &= M \left( \cos\varsigma \ch \left( \zeta_+ \nu \right) + \cos\zeta_+ \ch \left( \varsigma \nu\right) \right) \, , \\
    Y &= M \left( \sin\varsigma \sh \left( \zeta_+ \nu \right) + \sin\zeta_+ \sh \left( \varsigma \nu\right) \right) \, ,
\end{align}
\end{subequations}
with
\begin{equation}
    M = \frac{4\cos\delta \sh (\pi\nu)}{\ch(4\alpha\nu) + \cos(4\alpha)} \, , 
\end{equation}
and
\begin{subequations}
    \begin{align}
    G_x &= \Sigma_+ \sin\alpha \ch(\alpha\nu) 
     \, , \,\quad
    G_x^\dagger = \Sigma_+^\dagger \sin\alpha \sh(\alpha\nu) \, , \\
    G_y &= \Sigma_- \cos\alpha \ch(\alpha\nu)  \, , \quad
    G_y^\dagger = \Sigma_-^\dagger \cos\alpha \sh(\alpha\nu) \, , 
\end{align}
\end{subequations}
with
\begin{subequations}
    \begin{align}
    \Sigma_\pm &= 1-\ch(2\alpha \nu) \pm \cos(2\alpha) \, , \\
     \Sigma_\pm^\dagger &=  1+\ch(2\alpha \nu) \pm \cos(2\alpha) \, .
\end{align}
\end{subequations}
In addition, 
\begin{subequations}
    \begin{align}
    H_x &= \sin\alpha \, \big( \cos\eta \sh (\beta_+\nu) + P_+ \cos \xi_+ \big) \, , \\[3pt]
    H_y &= \cos\alpha \, \big( \cos\eta \sh (\beta_+\nu) + P_- \cos \xi_+ \big) \, , \\[3pt]
    H_x^\dagger &= \sin\alpha \left( \cos\eta \ch (\beta_+\nu) + P_+^\dagger \cos \xi_+ \right) \, , \\
    H_y^\dagger &= \cos\alpha \left( \cos\eta \ch (\beta_+\nu) + P_-^\dagger \cos \xi_+ \right) \, ,
\end{align}
\end{subequations}
where
\begin{subequations}
    \begin{align}
    P_\pm &= 2\sh(\pi\nu) \ch (\hat{\zeta}_- \nu) - \left( 1\pm 2\cos(2\alpha) \right) \sh(\beta_-\nu) , \\
    P_\pm^\dagger &= 2\sh(\pi\nu) \sh (\hat{\zeta}_- \nu) + \left( 1\pm 2\cos(2\alpha) \right) \ch(\beta_-\nu) .
\end{align}
\end{subequations}
Moreover, 
\begin{subequations}
\begin{align}
    \begin{pmatrix}
        K_x \\
        K_x^\dagger
    \end{pmatrix}
    &= 
    \begin{pmatrix}
        Q_1 \\
        Q_1^\dagger
    \end{pmatrix}
    \cos\tau -
    \begin{pmatrix}
        Q_2 \\
        Q_2^\dagger
    \end{pmatrix}
    \sh(\pi\nu) \cos\hat{\xi}_- \, , \\[3pt]
     \begin{pmatrix}
        K_y \\
        K_y^\dagger
    \end{pmatrix}
    &= 
    \begin{pmatrix}
        Q_1 \\
        Q_1^\dagger
    \end{pmatrix}
    \sin\tau -
    \begin{pmatrix}
        Q_2 \\
        Q_2^\dagger
    \end{pmatrix}
    \sh(\pi\nu) \sin\hat{\xi}_- \, ,
\end{align}
\end{subequations}
where
\begin{subequations}
    \begin{align}
    Q_1 &=  \sh \left( \overline{\alpha} \nu\right) \sh(\beta\nu) - \sh(\pi\nu) \sh( \hat{\zeta}_- \nu ) \, , \\
    Q_1^\dagger &=  \sh \left( \overline{\alpha} \nu \right) \ch(\beta\nu) - \sh(\pi\nu) \ch( \hat{\zeta}_- \nu ) \, ,
\end{align}
\end{subequations}
and 
\begin{equation}
     Q_2 = \sh(\beta\nu) + \sh (\hat{\zeta}_+ \nu) \, , \quad
    Q_2^\dagger = \ch(\beta\nu) - \ch (\hat{\zeta}_+ \nu) \, . 
\end{equation}

\section{Green's functions in real space}
\label{sec:green_real}

We apply the inverse FKL-transform, as given by Eq.~\eqref{eq:inv_FKL}, to express the displacement field in terms of integrals over the radial and axial wavenumbers. The integration with respect to~$k$ can be evaluated analytically, reducing the final expressions to a single improper integral over~$\nu$, which can be handled numerically.
Since $\widetilde{\phi}_j$ are even functions of~$k$, $j\in\{ x,y,w \}$, the inverse complex Fourier transform can be expressed in terms of the inverse cosine Fourier transform.

We define the operator 
\begin{equation}
    \Psi_j (\theta, \nu) = 
    \psi_j(\theta,\nu)
    + \varphi_j(\theta,\nu) \,
    \rho \, \frac{\partial}{\partial \rho} \, , 
\end{equation}
where
\begin{subequations}
    \begin{align}
    \psi_j (\theta, \nu) &= q_\perp \left(
\Lambda_j \sh(\theta\nu) + \Lambda_j^\dagger \ch(\theta\nu) \right) , \\
    \varphi_j (\theta, \nu) &= q_\perp \left(
\Omega_j \sh(\theta\nu) + \Omega_j^\dagger \ch(\theta\nu) \right) ,
\end{align}
\end{subequations}
and remark that $\varphi_w = 0$.
The solution is represented in the form
\begin{equation}
    \phi_j (r,\theta, z) = \int_0^\infty \Psi_j (\theta, \nu) \, 
    \mathcal{K}_{i\nu}(r,z) \, \mathrm{d}\nu \, 
    \label{eq:phi_j_final}
\end{equation}
for $j \in \{x,y,w\}$, where the kernel functions are defined as
\begin{align}
    \mathcal{K}_{i\nu} (r,z) = \left(\tfrac{2}{\pi}\right)^2  
    \int_{0}^\infty \cos(kz) K_{i\nu}(k\rho) K_{i\nu} (kr) \, \mathrm{d}k  \, . \label{eq:K_inu_def}
\end{align}
This improper integral converges. Its value is available in classical textbooks as
\begin{equation}
    \mathcal{K}_{i\nu} (r,z) = 
    \left( \rho r \right)^{-\frac{1}{2}}
    P_{i\nu-\frac{1}{2}} (\xi) \sch (\pi\nu) \, .
    \label{eq:Kinu}
\end{equation}
For instance, we refer to the comprehensive table of integrals by Gradshteyn and Ryzhik~\cite[p.~719]{gradshteyn2014table} Eq.~ET I 50~(51) or Prudnikov \textit{et al.}~\cite[p.~390]{prudnikov1992integrals} Eq.~2.16.36~(2).
Here, \( P_n \) stands for the Legendre function of the first kind of degree~$n$, and
\begin{equation}
    \xi = \frac{\rho^2+r^2+z^2}{2\rho r} .
\end{equation}

We have now formulated the unknown harmonic functions as integrals over the radial wavenumber \( \nu \), as shown in Eq.~\eqref{eq:phi_j_final}. 
While the improper integrals in these expressions could, in principle, be evaluated analytically using the residue theorem, we choose numerical integration for simplicity. 
By introducing the change of variable \( u = \nu / (\nu + 1) \), which gives \( \mathrm{d}\nu = \mathrm{d}u / (1 - u)^2 \), the infinite integrals are transformed into well-behaved integrals over the interval \([0,1]\), making them suitable for standard numerical integration techniques.

Finally, the displacement field results from Eqs.~\eqref{eq:displacement_field} as
\begin{align}
    u_j (\bm{x}) 
    &= u_j^\infty (\bm{x}) 
    + \int_0^\infty \mathscr{D}_j \, 
    \mathcal{K}_{i\nu} (r,z)
    \, \mathrm{d}\nu \, ,  
\end{align}
for $j \in \{r, \theta, z\}$, 
where the elements of the operator \( \mathscr{D}_j \) are defined as
\begin{subequations}
    \begin{align}
    \mathscr{D}_r &= \Psi_r \left( r\, \frac{\partial}{\partial r} - \sigma \right) 
    + \Psi_w \, \frac{\partial}{\partial r} \, ,\\
    \mathscr{D}_\theta &= \frac{\partial\Psi_r}{\partial \theta} - (\sigma+1) \Psi_\theta
    + \frac{1}{r} \frac{\partial\Psi_w}{\partial\theta} \, , \\
    \mathscr{D}_z &= r \Psi_r \, \frac{\partial}{\partial z} + \Psi_w \, \frac{\partial}{\partial z} \, ,
\end{align}
\end{subequations}
with the radial component $\Psi_r = \cos\theta \, \Psi_x + \sin\theta \, \Psi_y$ and azimuthal component $\Psi_\theta = \cos\theta \, \Psi_y - \sin\theta \, \Psi_x$.

\begin{figure}
    \centering
    \includegraphics[width=\linewidth]{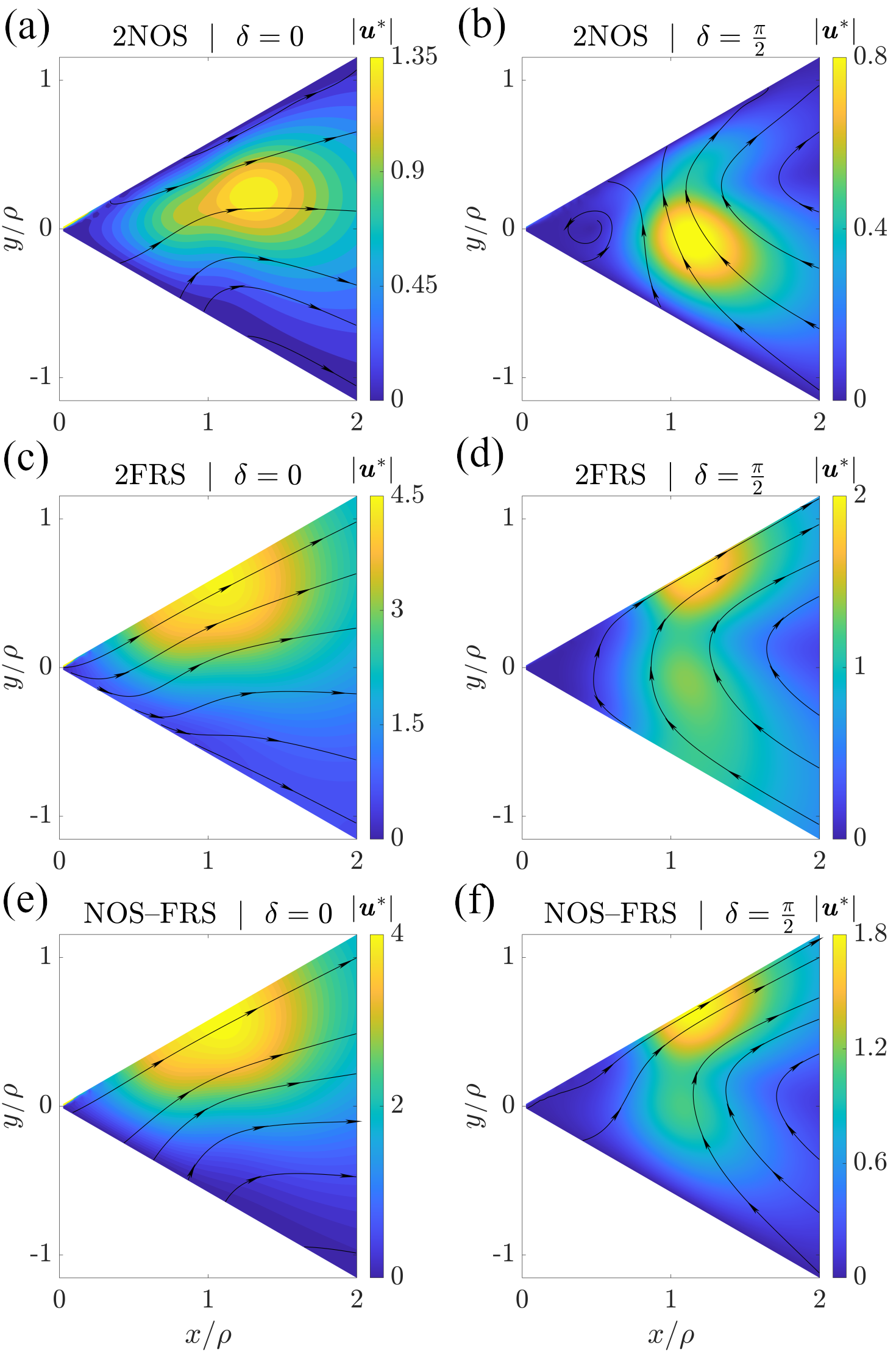}
    \caption{Scaled displacement field, $\bm{u}^* = \bm{u} / \left( F_\perp / \left( 16\pi\mu \right) \right)$ in the radial-azimuthal plane at $z/\rho = 1/2$ as induced by a point force acting perpendicular to the edge in the $x$-$y$-plane at $z=0$. The force points into radial direction ($\delta = 0$) in  (a), (c), and~(e). It is azimuthally directed ($\delta = \pi/2$) in (b), (d), and (f).
    A semi-opening angle of the wedge of $\alpha = \pi/6$ is considered, while the singularity is positioned at $\beta = \alpha/2$. We here set $\sigma = 2$. Concerning the different boundary conditions, we address 2NOS in (a) and~(b), 2FRS in (c) and~(d), and NOS--FRS in~(e) and~(f).}
    \label{fig:RThe}
\end{figure}

\begin{figure}
    \centering
    \includegraphics[width=\linewidth]{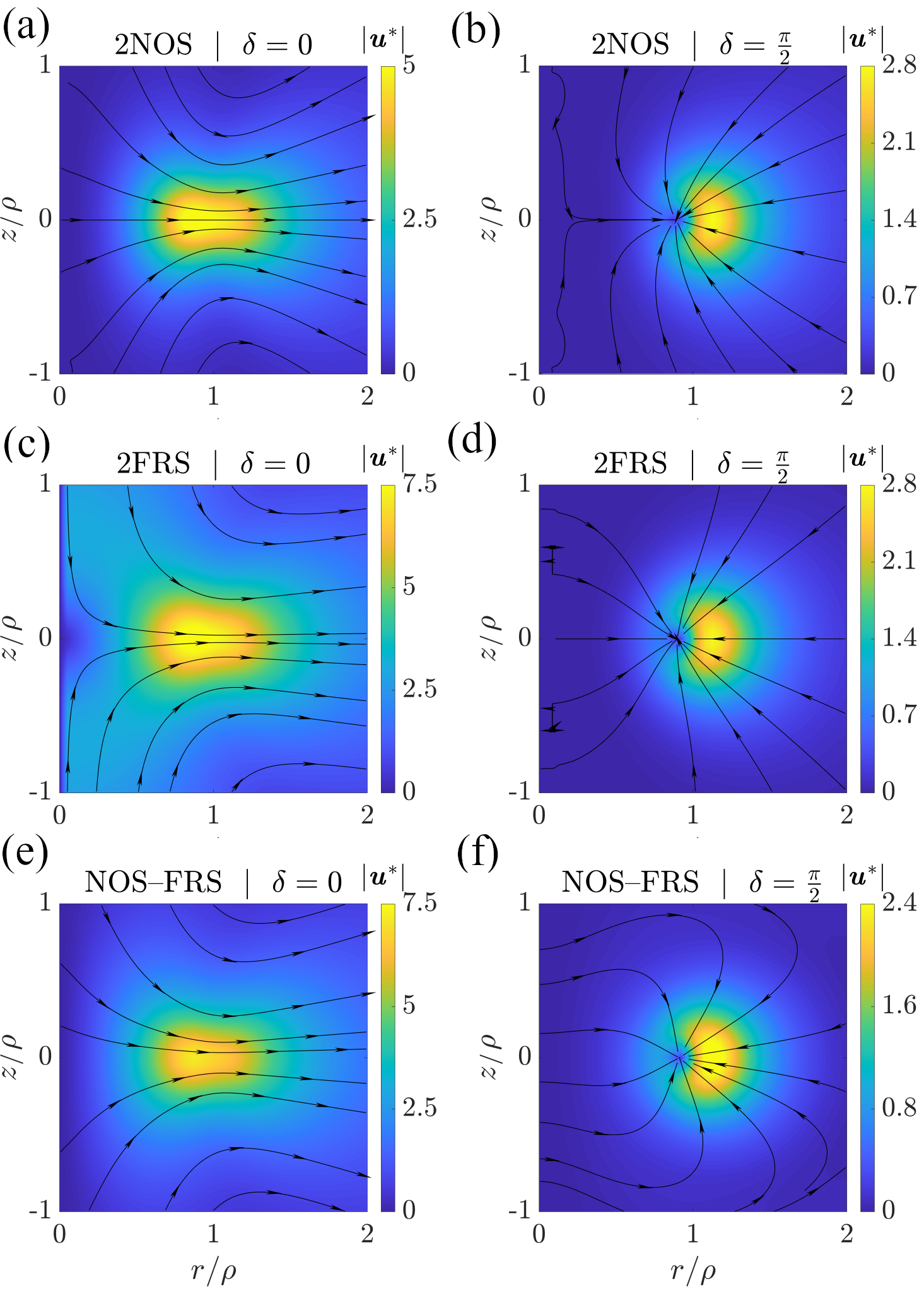}
    \caption{Displacement fields in the radial-axial plane at $\theta = 0$ ($x$-$z$-plane) for the same geometries and parameters as in Fig.~\ref{fig:RThe}.  
    Again, the displacement fields are rescaled to $\bm{u}^* = \bm{u} / \left( F_\perp / \left( 16\pi\mu \right) \right)$.}
    \label{fig:RZ}
\end{figure}

In Fig.~\ref{fig:RThe}, we present an example of the scaled displacement field in the radial-azimuthal plane at $z/\rho = 1/2$ for $\alpha = \pi/6$, with the point force positioned at the polar angle $\beta = \alpha/2$. The material is assumed to be linearly elastic with $\sigma = 2$.  
Results are shown for three different boundary conditions: 2NOS in (a) and (b), 2FRS in (c) and (d), and NOS--FRS in~(e) and (f). We apply the point force normal to the edge along the radial direction with $\delta = 0$ in (a), (c), and~(e). Conversely, it is applied along the azimuthal direction with $\delta = \pi/2$ in (b), (d), and (f). Numerical integration is performed using $N = 100$ collocation points, ensuring a well-resolved displacement field.

As may be expected, the no-slip conditions tend to suppress the magnitude of displacements. This becomes obvious when we consider the maximum amplitudes of displacement. They tend to increase the more we turn from no-slip to free-slip boundary conditions, that is, from 2NOS via NOS--FRS to 2FRS, see the numbers on the color bars in Fig.~\ref{fig:RThe}.   
Another visible effect of the boundaries is that the location of maximum displacement is shifted. 
While no-slip surfaces tend to push the locations of maximum displacement away from the surfaces of the wedge, free-slip surfaces tend to attract them. 
Interestingly, in the case of a purely azimuthally oriented force for 2NOS boundary conditions in Fig.~\ref{fig:RThe}(b), we observe a swirling structure of the displacement field close to the edge in the center of the wedge.
Yet, we remark that the magnitude of displacements is relatively low in this area. 

Considering the same geometries and the same parameters as in the panels of Fig.~\ref{fig:RThe}, we also address the displacement fields in the radial-axial plane. 
Corresponding displacement fields are presented in Fig.~\ref{fig:RZ}, showing the radial-axial plane at $\theta = 0$, that is, the $x$-$z$-plane. 
In this plane, there is no comparable shift of maxima in displacement fields as in Fig.~\ref{fig:RThe}. 

Finally, it is worth noting a limitation of the formalism used in this study.
While solutions to the NOS–FRS problem can be obtained in FKL space, it's important to note that the inverse-transformed solution is only valid for $\alpha \in (0, \pi/4]$—that is, it is restricted to wedge openings up to a right angle.

\section{Exact solutions in the limit of one planar boundary}
\label{sec:planar_boundary}

The semi-analytical results presented in Sec.~\ref{sec:green_real} for 2NOS and 2FRS boundary conditions are valid in the interval $\alpha \in (0, \pi/2]$. Therefore, it is possible to recover known results in the limit of one planar boundary. For this purpose, we let the semi-opening angle of the wedge approach $\pi/2$.

For the case of 2NOS, we obtain
\begin{subequations}
    \begin{align}
    \psi_w &= -\rho \, q_\perp \cos\delta \ch ( (\theta+\beta)\nu) \, , \\[3pt]
    \psi_x &= q_\perp \bigg( \cos(\beta+\delta)\ch((\theta+\beta)\nu) \notag \\
    &\quad-\frac{2}{\sigma} \, \cos\beta\sin\delta 
    \, \nu \sh((\theta+\beta) \nu)
    \bigg) , \\
    \psi_y &= q_\perp \sin(\beta+\delta) \ch((\theta+\beta)\nu) \, ,
\end{align}
\end{subequations}
together with
\begin{equation}
    \varphi_x = -\frac{2}{\sigma} \, q_\perp \cos\beta\cos\delta 
    \ch((\theta + \beta)\nu) \, , 
\end{equation}
and $\varphi_y = \varphi_w = 0$.
In contrast to that, for the situation of 2FRS, the coefficients in the planar boundary limit become
\begin{subequations}
    \begin{align}
    \psi_w &= \rho\, q_\perp \cos\delta \ch ((\theta+\beta)\nu) \, , \\
    \psi_x &= q_\perp \cos(\beta+\delta) \ch ((\theta+\beta)\nu) \, , \\
    \psi_y &= -q_\perp \sin(\beta+\delta) \ch ((\theta+\beta)\nu) \, , 
\end{align}
\end{subequations}
and $\varphi_j = 0$ for $j \in \{x,y,w\}$.

Therefore, in the limit of one planar boundary, the solution under 2NOS conditions can be written as
\begin{subequations}
    \begin{align}
    \phi_w &= -\rho \, q_\perp \cos\delta \,
    \mathscr{S}(\xi, \theta+\beta) \, , \\[3pt]
    \phi_x &= q_\perp
    \big( \cos(\beta+\delta) \, \mathscr{S} (\xi, \theta+\beta)
    - \cos\beta \, \mathscr{Q}  \big)  , \\
    \phi_y &= q_\perp
    \sin(\beta+\delta) \, \mathscr{S}(\xi, \theta+\beta) \, , 
\end{align}
\end{subequations}
with the differential operator 
\begin{equation}
    \mathscr{Q} 
    = \left. \frac{2}{\sigma} \left( \sin\delta \, \frac{\partial}{\partial a} + \cos\delta \, \rho \, \frac{\partial}{\partial \rho} \right) \mathscr{S}(\xi, a) \right|_{a=\theta+\beta} \, .
\end{equation}
Here, we have defined the improper integral
\begin{equation}
    \mathscr{S} (\xi, a) = 
    \left( \rho r\right)^{-\frac{1}{2}}
    \int_0^\infty \ch(a\nu) \sch(\pi\nu) \, P_{i\nu-\frac{1}{2}} (\xi) \, \mathrm{d}\nu \, , 
\end{equation}
which converges for $a \in [-\pi, \pi]$ and can be evaluated analytically as
\begin{equation}
     \mathscr{S} (\xi, a) = 
     \left( 2 \rho r \left( \xi + \cos a \right) \right)^{-\frac{1}{2}} \, .
\end{equation}
Similarly, for 2FRS conditions, we obtain
\begin{subequations}
    \begin{align}
    \phi_w &= \rho\, q_\perp \cos\delta \, \mathscr{S}(\xi, \theta+ \beta) \, , \\
    \phi_x &= q_\perp \cos(\beta+\delta) \, \mathscr{S}(\xi, \theta+ \beta) \, , \\
    \phi_y &= - q_\perp \sin(\beta+\delta) \, \mathscr{S}(\xi, \theta+ \beta) \, .
\end{align}
\end{subequations}

From there, the expressions for the harmonic functions can be simplified to
\begin{subequations}
    \begin{align}
    \phi_w &= -q_\perp \, \frac{\rho \cos\delta}{ \overline{s} } \, , \\
    \phi_x &= q_\perp \bigg( \frac{\cos(\beta+\delta)}{\overline{s}} + 
    \frac{2\rho \cos\beta}{\sigma \overline{s}^3} \notag \\
    &\quad\times \left( r\cos(\theta+\beta+\delta)+\rho \cos\delta \right)
    \bigg), \\
    \phi_y &= q_\perp \, \frac{\sin(\beta+\delta)}{ \overline{s} } \, ,
\end{align}
\end{subequations}
for 2NOS and to
\begin{subequations}
    \begin{align}
    \phi_w &= q_\perp \, \frac{\rho\cos\delta}{\overline{s}} \, , \\
    \phi_x &= q_\perp \, \frac{\cos(\beta+\delta)}{\overline{s}} \, , \\
    \phi_y &= -q_\perp \, \frac{\sin(\beta+\delta)}{\overline{s}} \, , 
\end{align}
\end{subequations}
for 2FRS. In these expressions
\begin{equation}
    \overline{s} = \left( r^2+\rho^2 + 2\rho r \cos(\theta+\beta) + z^2 \right)^{\frac{1}{2}} \, , 
\end{equation}
denotes the distance from the image system of the point force that is introduced when directly investigating the situation of a planar boundary.
It is straightforward to verify numerically that the resulting displacement fields are in good agreement with previous expressions for both no-slip and free-slip boundary conditions~\cite{menzel2017force}.

\section{Conclusions}
\label{sec:concl}

In summary, we present the derivation of the Green's functions for the deformation of a linearly elastic, homogeneous, isotropic, possibly compressible solid when enclosed by a wedge-shaped confinement bounded by a straight edge where its two surfaces meet. On these surfaces, either no-slip or free-slip conditions apply, while we also address the mixed situation of one surface showing no-slip and the other one featuring free-slip conditions. 

Our results for the linearly elastic solid carry over to the situation of low-Reynolds-number flows of viscous, homogeneous, isotropic, incompressible fluids, if we identify the displacement field with the flow field, the shear modulus with the shear viscosity, and set the Poisson ratio to one half. Overall, our solutions quantify the displacement or flow fields in response to a constant force applied at one point to the material within the wedge. 

In a previous work, we found the associated Green's functions for the situation of the applied force pointing into a direction parallel to the edge of the wedge. Contrariwise, we now present the solution for the force oriented perpendicular to the edge. In combination, these two cases form the general solution for the wedge-shaped geometry because of the linearity of the underlying equations. They can simply be superimposed. 
Likewise, the solution for a general, continuous force density field is obtained by spatial convolution with our solution.

Our resulting expressions can be used to study abundant situations such as sedimenting colloidal particles in suspensions maintained in a vessel \cite{jones1999sedimentation}, propulsion of active microswimmers near edges \cite{das2015boundaries}, or the excitation of soft elastic composite actuators \cite{chung2021magnetically} near a kink in their confinement, if the no-slip or free-slip boundary conditions apply. This approach can replace more extensive simulations using, for instance, multiparticle collision dynamics \cite{gompper2009multi, goh2025alignment} or finite-element methods \cite{kalina2023multiscale} in these specific geometries. Possibly, they can be combined with simulation methods to speed up corresponding evaluations or serve as input for optimization routines \cite{fischer2024maximized}. 
It would thus be valuable to explore related systems in exchange with advanced computational methods to address real-world applications. This is an avenue that we may consider pursuing in a future combined numerical and theoretical study.
Possible worthwhile further extensions may concern conditions of partial slip \cite{lauga2005brownian} or stress-free boundary conditions. Particularly for elastic solids under unloaded conditions, the latter would represent another relevant situation.

\section*{Data availability}

The data and numerical codes that support the findings of this study are available from the corresponding author upon reasonable request.

\section*{Author contribution}

A.D.M.I.\ and A.M.M.\ designed the research and wrote the paper. A.D.M.I.\ performed the analytical calculations and created the figures. 
Both authors reviewed, edited, and approved the final version of the manuscript.

\begin{acknowledgments}
A.M.M.\ thanks the Deutsche Forschungsgemeinschaft (German Research Foundation, DFG) for support through the Heisenberg grant ME 3571/4-1. 
\end{acknowledgments}

\appendix*

\section{FKL-transform of the Laplacian}

The Laplacian of a function $f$ in cylindrical coordinates takes the form
\begin{equation}
    \bm{\nabla}^2 f = \frac{1}{r} \frac{\partial}{\partial r}
    \left( r \, \frac{ \partial f }{\partial r} \right)
    + \frac{1}{r^2} \frac{\partial^2 f}{\partial \theta^2}
    + \frac{\partial^2 f}{\partial z^2} \, .
    \label{eq:Laplcian}
\end{equation}
It is simpler to compute the FKL-transform of $r^2 \bm{\nabla}^2 f$.
For the radial component of the Laplacian, we perform integration by parts twice.
In both instances, the boundary terms arising from partial integration vanish due to the applied boundary conditions.
In doing so, we make use of the fact that
\begin{equation}
    \frac{\partial}{\partial r}
    \left( r\, \frac{\partial}{\partial r} K_{i\nu} (|k|r) \right)
    = \frac{1}{r}\left( k^2 r^2-\nu^2 \right) K_{i\nu} (|k|r) \, 
\end{equation}
along with
\begin{equation}
   \mathscr{T}_{i\nu} \left\{ \frac{\partial^2 f}{\partial z^2} \right\} = -k^2 \mathscr{F}_{i\nu} \{f\} \, , \label{eq:app.diff_f_zz}
\end{equation}
to obtain
\begin{equation}
    \mathscr{T}_{i\nu} \left\{ 
    r\, \frac{\partial f}{\partial r}
    + r^2 \left( \frac{\partial^2 f}{\partial r^2} + \frac{\partial^2 f}{\partial z^2} \right) \right\}
    = -\nu^2 \mathscr{T}_{i\nu} \{f\} \, .
    \label{eq:app.Lap_axisymm}
\end{equation}
This final results reads
\begin{equation}
    \mathscr{T}_{i\nu} \left\{ r^2 \bm{\nabla}^2 f \right\}
    = \left( \frac{\partial^2}{\partial\theta^2}-\nu^2 \right) \mathscr{T}_{i\nu} \{f\} \, .
\end{equation}
This is the same as Eq.~\eqref{eq:ODE-2nd-order}.

%


\end{document}